# A Dynamic Power Management Schema for Multi-Tier Data Centers

Aryan Azimzadeh
Department of Computer Science
East Carolina University
Greenville, NC 27858

Nasseh Tabrizi
Department of Computer Science
East Carolina University
Greenville, NC 27858

*Abstract*— **An issue of great concern as it relates to global warming is power consumption and efficient use of computers especially in large data centers. Data centers have an important role in IT infrastructures because of their huge power consumption.**

**This thesis explores the sleep state of data centers under specific conditions such as setup time and get optimal number of servers. Moreover, their potential to greatly increase energy efficiency in data centers. This research uses a dynamic power management policy based on a mathematical model. The methodology is based on how it can get to the optimal number of servers required in each tier while increasing servers' setup time after sleep mode to reduce the power consumption. The reactive approach is used to prove the results are valid and energy efficient. It calculates the average power consumption of each server under specific sleep mode and setup time. Then average power consumption is used to get the Normalized-Performance-Per watt to evaluate the power efficient methodology. The results indicate that the proposed schema is beneficial for data centers with high setup time.**

I. INTRODUCTION

Data centers are already an essential part of internet services and have a growing role in several industries and businesses beyond the computer industry. In fact, all networking activity relies on data centers. Every time we search or use an application on our desktops, smartphone, or any platform of computers, data centers are required.

When considering data centers in industry and at a high level, it means that a large amount of servers is necessary, which contributes a massive number of computational resources, storing data to servers. Considering the massive collection of servers, energy consumption should also be understood. The energy consumption in data centers is one of the biggest factor in data centers expenses [1, 2]. Energy consumption is now the major issue for data centers. The results of a study show that data centers consume about 2.8% of the total electricity in the USA [3], Moreover, these centers' energy consumption represents about 3% of global energy use [4]. The main consumers of power within data centers are cooling systems and computing resources. Researchers estimate that cooling is around 30% of total energy consumption [5].

In response to many concerns about growing power consumption in data centers, many businesses are attempting a new strategy called green computing. The concept of green computing is to save energy, improve efficacy, and achieve environmental protection and energy saving [6]. Recent advances in energy efficiency have yielded huge improvements in both desktop and server computer technology. At the same time, industries are faced with contributing problems that relates to computer system, including the energy consumption, exhausted emissions, building resources, high maintenance costs, global warming, and high water enterprise [7, 8]. Green computing can reduce the energy consumption of computer systems, improve their operational efficiency of emissions, and increase recycling efficiency, which could promote environmental protection and conservation of energy [9].

Today's data centers are mostly working under AlwaysOn Policy, which wastes a lot of power during periods of lower loads [10]. Researchers have proposed various solutions to reduce energy consumption by optimizing servers with a sleep mode. A servers' setup time is one of the recent challenges in dynamic power management. Some researchers believe that it is not efficient to have have a high setup time, but this research will explain that its not always true. Current approaches to managing the server sleep state include the Predictive approach, the Reactive approach, hybrid approaches, and dynamic provisioning approaches in operations research amongst others.

The primary objective of this thesis is to compare the hybrid and reactive approaches, and secondly, to prove that under specific circumstances the combination of these two approaches can create a new approach in green data centers.

The next chapter will overview a related work and explain the challenges that researchers are facing in dynamic power management with server systems. The AlwaysOn policy will then be explained in Chapter III. Chapter IV will explain the theoretical methodology, and Chapter V will validate the methodology through the results produced. Finally, a summary of this thesis will be presented in Chapter VI.

II. RELATED WORKS

F In this section, the related work in data center dynamic power management is discussed. The prior work in different aspects of dynamic power management will be explained and highlighted. In order to demonstrate the resulting method as superior, one must explore related systems and analyze the tradeoffs in various approaches.

When using power management, in order to improve the energy efficiency of data centers, three techniques are commonly employed: selected servers shutdown, frequency and voltage provisioning, and dynamic power management [11]. There are two different kinds of dynamic power management: predictive and reactive [12]. The predictive approaches will envision the future request rate using previous data in order to recognize when the servers must be turned on [12]. On the other hand, the reactive approaches will react to the request immediately by turning the servers on or off. There is also another branch, which is the hybrid approach. The hybrid approach includes both predictive and reactive methods.

## A. Predictive Approaches

One of the approaches in this area is to use different types of predictive policies, such as exponentially weighted average, moving window average, and linear regression in order to predict the future request rate and add or remove servers based on the results. The authors determined that using moving window policies and linear regression enabled the best results for the workload traces that they considered. Thus, this methodology provides a means to more efficient power consumption than static approaches have in the past [13].

In another approach, researchers used auto regression policy to predict the request rate for specific arrival patterns and used the result of this calculation to determine the threshold policies that were able to trigger the servers on and off. Their dynamic power management policy is energy efficient for periodic request rates repeating on a daily basis [14].

## B. Reactive Approaches

In [15] the authors used a theoretical method as a control in order to manage resources to applications in a multi-tier data center. They used specific queuing theory to predict response time and allocated resources based on the estimated response time and power consumption.

Another approach used by the author involved using a reactive feedback mechanism to monitor a multi-tier web application. The author evaluated CPU utilization and response time and changed the number of severs based on these calculations, making the point that using multiple sleep states in servers could have significant improvement in energy savings [16].

There are some other approaches which mostly study modeling and dynamic provisioning on the performance side of multi-tier, and the approaches barely focus on power consumption [17, 18].

In [10] the authors proposed two different approaches called: Reactive and SoftReactive. Their results are achieved under different traces. Sleep states appreciate dynamic power management. They describe certain types of traces, evaluate them, and figure out which is the best match for sleep states. Reactive approach responds to changes in requests and loads by turning the servers to sleep mode and waking them back up when the load increases. There has been big concern for the Reactive policy; In reactive policy the servers go off so quickly when not needed, but when the loads rise, it takes time for servers to come back on again. So, to cure this problem, they introduce another policy called SoftReactive. In SoftReactive approach, the server goes to idle mode for a short time before it turns off. This delay in transition gives the opportunity for the server to wait for possible arrival load. If the server gets requested during the delay time, then the server goes back to the regular mode. The researchers set timers for each server to turn off, and the idea prevents the mistake of turning the server on at the wrong time. The problem raises in this methodology when the researchers put too many servers in the idle mode. To solve this issue, they introduced a routing plan, which distributes jobs onto the low amount servers, so the unneeded servers will go into sleep mode.

## C. Hybrid Approaches

Hybrid approach includes both predictive and reactive approaches. Predictive methods are used in long-term workload trends, and reactive methods are used in short-term unpredictable trends [12].

In [19], the authors first used the reactive method for unpredictable trends in request rate and later used the predictive method for long term trends in request rate. Separately, the authors proposed a solution called PowerNap. The authors' system has a way to switch its state from high performance to low power (sleep mode) and vice versa to respond to the rapid server loads. Using this methodology, the authors were able to put the servers in sleep mode long before the servers go into idle mode, so they are actually replacing the low server utilization periods with an energy efficient sleep mode [20].

In [11], the authors introduce new methodology consisting of multiple approaches. They use dynamic provisioning, frequency scaling, and dynamic power management methods to make multi-tier data centers more energy efficient. They propose two algorithms; one focuses on the optimal number of servers by dynamically provisioning them, and the other algorithm, mostly focuses on the CPU speed and the duration of sleep states for each server.

Unfortunately, thus far, hybrid approaches have had problems predicting workloads and reactive approaches, but this research will attempt to overcome part of this problem by using some aspect of hybrid and reactive approaches together.

III. ALWAYSON POLICY

AlwaysOn is a static power management policy, which most of the industries nowadays are using. The policy has a constant number of active front-end servers at all times. To figure out how many servers this policy uses, the amount of request rates that each front-end server can handle must be obsevred [10]. This is the critical point, when the $95^{th}$ percentile of certain threshold will be implemented.

This policy is designed to meet peak request rate, but it does not have the ability to envision when peak request rate occurs. The average power consumption for the AlwaysOn policy is always high. Moreover, the $95^{th}$ percentile of response time and average power consumption under AlwaysOn are unchanged in a favor of sleep states. That is why the AlwaysON policy was chosen to compare the approach established by this research and the reactive approach.

## IV. METHODOLOGY

In this section, the methodology based on two previous approaches using dynamic power management will be described. The goal of this research is to point out the fact that it can be proven, under specific conditions, that two different methodologies can be combined to get a third, green approach, in the field of dynamic power management in data centers.

For the methodology, it was assumed that we have one front-end load generator and one front-end load balancer, which distributes request from the load balancer to expected application servers. The load balancer is also being responsible for turning the application server to sleep mode and waking them up. There are also several memcached servers to fetch data required to service the requests [10]. The point is, the researcher applied all power management techniques on the front-end application server side. So whenever the number of severs in this research is mentioned in this research, it refers to the front-end application server.

The (1) [11] has been used to obtain optimal energy consumption. Then (2) is used to convert the energy to power. The goal of this conversion is we used $T_{setup}$ based on the 95 percentile of customers response time. In our methodology we use TPC-W [21] based workload in multi-tier data center.

$$E = P(s, 1.0)[(T - t)(\rho(1 - k) + k) + tk'] \quad (1)$$

$$P = \frac{E}{T_{setup}} \quad (2)$$

By using the (1) we get to a point where $E = 250\ (Wh)$ is our optimal energy consumption. In (1), $P$ is the power consumption, $\rho$ represents the utilization of a system, $s$ represents the CPU speed, t represents sleep state duration and $T$ represents time interval length.

The parameters of interest include the average power consumption, setup time, response time and the number of active servers. Setup time is defined as the time that servers take to turn back on from sleep mode. Although, long setup times are not recommended commonly, we will show that if specific time slots are considered in our calculation in combination with the specific number of servers, it can be efficient to use long setup times.

Our consideration for hybrid aspects include the CPU speed and also how to get to the minimal number of servers. So we get the expected minimal number of servers by (3) [11]. The minimal number of servers that we get from (3) should meet the Service Level Agreement, SLA, and can help us to achieve good ratings in our power saving approach. The CPU utilization can be obtained by monitoring the supported tools by operation systems. Then we analyze the number of requests by a server in different time frames.

$$v_i = \frac{L_i + r_i}{T_{SLA} + \tau_i} \quad (3)$$

$$Front - end\ servers = \left\lceil \frac{r}{60} \right\rceil \quad (4)$$

In (3), $v_i$ represent the minimal number of servers in each tier, $L_i$ is number of queued for each tier, $r_i$ is number in incoming request for each tier, $T_{SLA}$ is the target response time and $\tau_i$ is estimated throughout tier $i$.

The results in (3) can be validated by the Algorithm 1 mentioned in [11]. In our approach, the number of servers that we got from (3) will be validated with the peak number of requests in (4) [10]. Each front end server can handle 60 $req/s$ [10]. This result is based on a $T_{95}$ threshold of 500. In this research we compare our results with the AlwaysOn policy and reactive policy. Note that in the real world, the request rate cannot be calculated in advance, but we assume the request rate in advance from the AlwaysOn policy.

We assume a peak request of 800 $req/s$ for specific benchmark dynamically over 30 minutes, and our dynamic power management scheme calculates the number of servers for each tier during the next time interval. Based on (4), $\left\lceil \frac{800}{60} \right\rceil =14$ servers for the AlwaysOn policy are needed at all times, but this number can vary in our schema.

With reactive policy, the servers react to the ongoing request rate and can adjust their capacity in real time. However, in our approach it has been said that reactive policy suffers from long setup times. We show that it can prove power efficient to use it in our way.

Our approach puts the servers in sleep mode if the actual number of servers are more than $\left\lceil \frac{r}{60} \right\rceil$ and if not the servers are called to come back from sleep. The other way that we put the servers in sleep mode is when there is delay in the incoming requests. But how can we prove that our approach is energy efficient and works with a high setup time and a limited number of servers?

In order to determine for how long, the servers are put in sleep mode and the response time for each request is estimated. (5) [11] is used to get to approximate response time in 30 minutes setup time. In this equation, $L_i$ is the number of requests, $n$ is the number disciplines that CPU need to process the request and $s$ is the CPU speed.

$$\frac{(L_i + 1)n}{s} \quad (5)$$

"Fig. 1" shows $T_{95}$ of response time for the TCP-W benchmark for each time slots. As shown below, the $T_{95}$ starts from $T$ = 12 minutes, so in our calculation we exclude time frames before $T$ = 12 minutes to meet The SLA limit of 2000ms. The Hybrid approach keeps the response time below 2000ms, thus making it easier to allocate the expected number of servers [11].

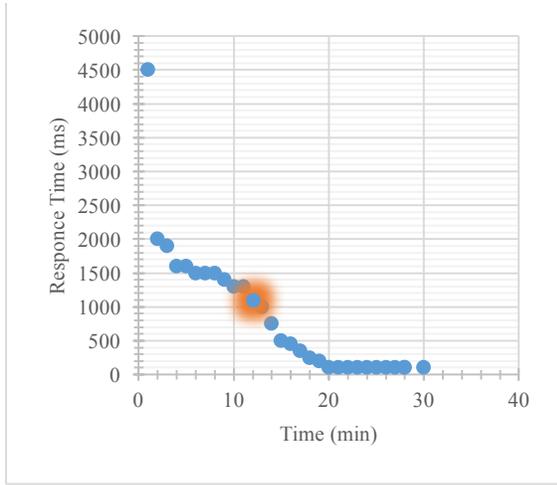

Fig. 1. Calculated $T_{95}$ response time

We also need to get the the average power consumption, $P_{avg}$, for our calculation. The approach was to clone the influence of using sleep state, by not sending the request to the server when it is marked for sleep and changing its power consumption by $P_{sleep}$. To prove our approach is energy efficient we use a metric call, Normalized Performance per Watt, NPPW, but before NPPW is calculated, another metric called Performance per Watt, PPW, is required. We need PPW for both Reactive and AlwaysOn policy [10].

*A. Average Power Consumption Calculation*

In this section explains how to calculate the $P_{avg}$ in our reactive approach. To get the $P_{avg}$ for specific setup time, the power consumption in that time must be calculated first. As mentioned in paragraph three of this chapter we lead to energy consumption of $250\ Wh$. $P_{avg}$ is different for various setup times when the $P_{sleep}$ is zero. The $P_{avg}$ is calculated based on setup times which starts from time slot 12 minute. Paragraph three also mentioned that $T_{setup} = 12\ minutes$ is the the start point of our $T_{95}$ of response time. First we calculate the $P_{avg}$ when the server $P_{sleep}$ is zero and then increase the setup time in this state to get power consumption. Although $T_{setup} = 12\ minutes$ is a start point of our $T_{95}$ of response time, we start the calculation from $T_{setup} = 15\ minutes$ because based on our calculation the power consumption before $T_{setup} = 15\ minutes$ is not efficient. After the first calculations of $P_{avg}$ for $P_{sleep} = 0$ then $P_{sleep}$ increases [10]. We predict $P_{avg}$ for a given $T_{setup}$ and $P_{sleep}$ by analyzing the results in [10]. Note that, all the results for $P_{avg}$ is in reactive mode. "Fig. 2" shows our results for $P_{avg}^{Reactive}$.

|  | $P_{sleep}$(watts) | | | |
|---|---|---|---|---|
|  | 0 | 28 | 56 | 84 |
| 15 | 1000 | 1279 | 1558 | 1837 |
| 16 | 937 | 1216 | 1495 | 1774 |
| 17 | 833 | 1161 | 1440 | 1719 |
| 18 | 883 | 1112 | 1391 | 1678 |
| 19 | 789 | 1068 | 1347 | 1620 |

Fig. 2. Results for our approach with respect to $P_{avg}$

As seen in "Fig. 2", $P_{avg}$ decreases when we increase the $T_{setup}$ at the same $P_{sleep}$. On the other hand, when $P_{sleep}$ increases, $P_{avg}$ increases at the constant level of the $T_{setup}$. For instance, for $P_{sleep} = 0$, when the $T_{setup}$ increases from 15 minutes to 19 minutes, $P_{avg}$ decrease from 1000 Watts to 789. On the other hand, for $T_{setup} = 19$ minutes, $P_{avg}$ boost form 789 Watts to 1620 watts. We will explain later in this chapter that low $P_{avg}$ is so beneficial for our system.

*B. Performance-Per-Watt Calculation*

PPW is extremely important to our calculations. Note that higher PPW is better to get to the improved energy efficiency. Equation (6) shows how to calculate the PPW. It also shows that for each specific $T_{setup}$, we get the same value of $T_{95}$ by increasing $P_{sleep}$. So the calculation has only five different $T_{95}$ values as we are considering five different $T_{setup}$. $T_{95}$ increases as $T_{setup}$ increases. "Fig. 3" shows the PPW calculation. The results for PPW show that by increasing $P_{sleep}$ at specific $T_{setup}$, PPW decrease and by contrast, when $P_{sleep}$ is constant, PPW increases by increasing $T_{setup}$. That is why when we have the maximum value of PPW when $T_{setup} = 19$ minutes and $P_{sleep} = 0$. Note that, PPW for AlwaysOn is unaffected by changes in $P_{sleep}$ and $T_{setup}$ and it has a constant value of $1.7 \cdot 10^{-6}\ (ms.watts)^{-1}$ [10]. We will use these values of PPW in the next chapter to get NPPW.

$$PPW = \frac{1}{P_{avg} \cdot T_{95}} \quad (6)$$

|  | $P_{sleep}$(watts) | | | |
|---|---|---|---|---|
|  | 0 | 28 | 56 | 84 |
| 15 | $2*10^{-6}$ | $1.5*10^{-6}$ | $1*10^{-6}$ | $1*10^{-6}$ |
| 16 | $2*10^{-6}$ | $1.8*10^{-6}$ | $1.5*10^{-6}$ | $1.3*10^{-6}$ |
| 17 | $3*10^{-6}$ | $2*10^{-6}$ | $2*10^{-6}$ | $1.3*10^{-6}$ |
| 18 | $5*10^{-6}$ | $4*10^{-6}$ | $3*10^{-6}$ | $2*10^{-6}$ |
| 19 | $6*10^{-6}$ | $4*10^{-6}$ | $4*10^{-6}$ | $3*10^{-6}$ |

Fig. 3. Results for our approach with respect to PPW

## V. RESULTS

We got PPW for various sleep states duration from previous chapter. Now these values are used to prove that not only are our results superior to the AlwaysOn policy, but they are also superior to a purely reactive approach [10]. To prove this we will need to get NPPW for all $T_{setup}$ and $P_{sleep}$. Equation (7) shows how to calculate NPPW by normalizing PPW for Reactive by PPW for AlwaysON [10].

$$NPPW = \frac{PPW^{Reactive}}{PPW^{AlwaysOn}} \quad (7)$$

When NPPW exceeds 1, it demonstrates that our approach is superior to AlwaysOn. This means that our result is more energy efficient. By the use of the optimal number of servers (approximately 60 servers) which is based on (3) and (4) we got to the results that are mentioned in "Fig. 3". The results from "Fig. 3" are used as $PPW^{Reactive}$ in (7) to calculate NPPW. As mentioned in previous chapter, for AlwaysOn Policy $PPW = 1.7 \cdot 10^{-6} \ (ms.watts)^{-1}$.

"Fig. 4" shows our result for NPPW for slowly varying traces. White regions demonstrate higher NPPW, where NPPW > 1 argue that our approach is superior to AlwaysOn Policy.

|  | $P_{sleep}$(watts) | | | |
|---|---|---|---|---|
| $T_{setup}$(min) | 0 | 28 | 56 | 84 |
| 15 | 1.17 | 0.89 | 0.59 | 0.59 |
| 16 | 1.17 | 1.06 | 0.89 | 0.76 |
| 17 | 1.76 | 1.17 | 1.17 | 0.94 |
| 18 | 2.9 | 2.35 | 1.76 | 1.17 |
| 19 | 3.53 | 2.35 | 2.35 | 1.76 |

Fig. 4. Normalized Performance-per-Watt (NPPW) under our approach

This figure shows that NPPW increases as $T_{setup}$ increases and $P_{sleep}$ decreases. As an illustration we have a maximum NPPW of 3.53 when the $T_{setup} = 19 \ minutes$ and $P_{sleep} = 0$.

We find that using sleep states under reactive and hybrid policy can provide demonstrable benefit in terms of NPPW for specific conditions. Using the specifications form calculation and the result of NPPW, we were able to achieve significant improvements in energy efficiency compared to previous schemas.

It is revealed that the evolution of our approach is superior to previous Reactive approach and AlwaysOn policy. "Fig. 5" shows that the results of our method compared to those previously acquired by other groups while scaling the number of servers up from 14 to 60, magnification increases NPPW. While not usually recommended, the results make our approach more desirable as compared to AlwaysOn and Reactive policies.

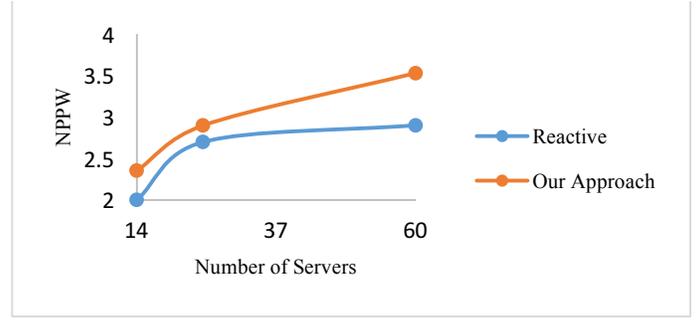

Fig. 5. Effect of scaling on NPPW and comparison of our results to Reactive

## VI. CONCLUSION

The new methodology was introduced to examine the benefit of sleep states with high server setup times. The methodology is the combination of reactive and hybrid approaches to get to our goal which is used different equations to find the minimal and optimal number of servers, then we validate those results with different equation related to request rates. Our methodology needed 95[th] percentile of response time, so we calculated this for our specific setup time. We used certain ranges of sleep states with different high setup times and proved that it can extremely boost Performance Per Watt, PPW. The PPW results shows that our results are much more superior to AlwaysOn. Then We calculated Normalized-Performance-Per watt and proved that our approach is also superior to previous Reactive approach under specific circumstances. Finally, we compared our result by increasing the number of serves with Reactive approach, our examination shows the effectiveness of sleep states when the number of servers increases. In particular, our results express that the proposed schema can reduce the power consumption by 48% relative to static provisioning and AlwaysOn policy.